# IBIC CHARACTERIZATION OF AN ION-BEAM-MICROMACHINED MULTI-ELECTRODE DIAMOND DETECTOR

J. Forneris[1], V. Grilj[2], M. Jakšić[2], A. Lo Giudice[1], P. Olivero[1], F. Picollo[1], N. Skukan[2], C. Verona[3], G. Verona-Rinati[3], E.Vittone[1#]

[1] Dipartimento di Fisca e Centro di Eccellenza NIS - Università di Torino; INFN - sez. Torino; CNISM - sez. Torino; via P. Giuria 1, 10125 Torino, Italy

[2] Ruđer Bošković Institute, Bijenicka 54, P.O. Box 180, 10002 Zagreb, Croatia

[3] Dipartimento di Ingegneria Industriale, Università di Roma "Tor Vergata", Via del Politecnico 1, 00133 Roma, Italy

**Keywords:** diamond detector, Ion Beam Induced Charge, Deep Ion Beam Lithography

**ABSTRACT**

Deep Ion Beam Lithography (DIBL) has been used for the direct writing of buried graphitic regions in monocrystalline diamond with micrometric resolution. Aiming at the development and the characterization of a fully ion-beam-micromachined solid state ionization chamber, a device with interdigitated electrodes was fabricated by using a 1.8 MeV $He^+$ ion microbeam scanning on a homoepitaxial, 40 μm thick, detector grade diamond sample grown by chemical vapour deposition (CVD). In order to evaluate the ionizing-radiation-detection performance of the device, charge collection efficiency (CCE) maps were extracted from Ion Beam Induced Charge (IBIC) measurements carried out by probing different arrangements of buried microelectrodes.

The analysis of the CCE maps allowed for an exhaustive evaluation of the detector features, in particular the individuation of the different role played by electrons and holes in the formation of the induced charge pulses.

---

[#] Correspondig author; Ettore Vittone; e-mail:ettore.vittone@unito.it.

Finally, a comparison of the performances of the detector with buried graphitic electrodes with those relevant to conventional metallic surface electrodes evidenced the formation of a dead layer overlying the buried electrodes as a result of the fabrication process.

# 1. INTRODUCTION

In recent papers [1,2] we described the fabrication of buried graphitic channels in monocrystalline diamond by Deep Ion Beam Lithography (DIBL).This technique consists in a selective damage of the crystal induced by MeV ion beams, which are focused down to a micrometer spot size and raster scanned on the sample along predefined linear patterns intersecting slowly-degrading metallic masks.

The damage induced by ions is localized mainly at their end of range, i.e. few micrometers below the surface; the regions which experience a vacancy density overcoming a critical level, usually referred as "graphitization threshold", convert to a graphitic phase upon thermal annealing; elsewhere, the diamond structure is recovered.

With this method, highly conductive (resistivity of the order of 1 mΩ·cm) graphitic channels can be realized in single crystal diamond; their length is limited by the elongation of the micro-ion beam scanning system (typically several hundreds of μm), their minimum width is given by the beam spot size and their formation depth is defined by the nuclear stopping range of the ions in diamond (typically few micrometer). Moreover, the presence of metallic masks, which modulate the nuclear stopping power of the ions determine the emersion of the channels' terminals at the surface, allowing the bonding of the graphitic channels to external electronic systems or the selective etching of the graphite for subsequent diamond micromachining processes [3].

The realization of highly conductive, optically opaque, chemically reactive graphitic channels embedded in a highly resistive, optically transparent and chemically inert diamond matrix is of potential interest in many sectors, e.g. for the realization of diamond 3D microstructures [4], microfluidic channels, innovative biosensors [5, 6?], IR emitters [7] or bolometers [8]. or An additional source of interest is given by the possibility of exploiting the DIBL to realize novel 3D architectures for ionizing radiation detection, which can contribute to increase the charge collection efficiency and enhance the radiation hardness of diamond detectors [9].

With the purpose of characterizing the carrier transport and recombination features of micromachined diamond detectors with buried electrodes, an analytical techniques able to map the charge collection efficiency at a micrometer level of resolution is needed, with a well defined probing depth profile suitable to analyze the entire region where pulse signals are formed.

In this paper it is shown that the Ion Beam Induced Charge [1,10] technique fulfils these requirements. CCE maps obtained by raster scanning 4 MeV He$^+$ ion micro-beams onto regions surrounding the buried graphitic electrodes provide valuable information not easily available otherwise on the electronic characteristics of the detector, concerning the electric field distribution, the different role played by electrons and holes in the induced charge signal and the influence of the residual damage induced by the DIBL process to the diamond structure.

## 2. EXPERIMENTAL

The sample under test (sample 1) consisted in an intrinsic single-crystal ~40 $\mu$m thick homoepitaxial diamond layer grown on a commercial 4×4×0.4 mm$^3$ high pressure high temperature (HPHT) Ib single crystal diamond substrate at the laboratories of Rome "Tor Vergata" University, using a Microwave Plasma Enhanced Chemical Vapour Deposition (MWPECVD) process [11].

Graphitic channels buried into the diamond sample were made by DIBL [1,2] at the AN2000 microbeam line of the INFN National Laboratories of Legnaro (I), using a 1.8 MeV He$^+$ ion microbeam (~10 $\mu$m spot size ) with an ion current of ~1 nA.The ion fluence was around 1.5·10$^{17}$ cm$^{-2}$. From a SRIM2011 simulation [12], these irradiation conditions are suitable to produce a vacancy density profile with a damage peak well above the graphitization threshold [13] and located at a depth of about 3 μm, as shown in Fig. 1.

The electrical continuity to the surface was ensured by the evaporation of variable thickness and slowly-degrading Cu masks onto the diamond surface before implantation, allowing for the modulation of the depth of the ion beam-induced damaged channels at their endpoints, es previously explained [1,2].

After ion implantation, Cu masks were removed from the surface. The sample was then annealed in vacuum at 1100 °C for 2 hours, in order to convert the highly-damaged regions located at the ion end of range to a graphitic phase while removing the structural sub-threshold damage introduced in the layer overlying the above-mentioned damaged region (to which we will refer as the "cap-layer" in the following) [1,2].

As a result of microfabrication process (see Fig. 2a), in the sample four parallel buried graphitic channels were fabricated, plus an additional channel orthogonal to them, each one being ~10 μm wide. The average spacing between channels was ~12 μm.

To connect the graphitic channels to external electronic circuits, their emerging endpoints were contacted with 80 nm thick Cr/Al circular layers (150 μm diameter), which were used as bonding pads.

Electrical measurements between any pair of electrodes showed currents below the detection limit of our electrometer (<1 pA at ±100V applied bias), demonstrating both the high resistivity of the diamond matrix embedding the buried channels and a surface leakage current below the detection limit.

Due to the epitaxial growth process and to the thickness of the HPHT substrate, the diamond sample was not equipped with a back electrode. Therefore, configuration and geometry of the electric field in the diamond bulk was entirely defined by the voltage applied at the buried electrodes.

A nominally identical intrinsic diamond sample (sample 2) was grown using the same CVD process, in order to provide a comparison between buried graphitic channels and standard ohmic surface metallic contacts. After CVD growth, Ti/Pt/Au (50/20/50 nm) finger contacts were patterned by a standard lift-off photo-lithographic technique and by thermal evaporation on the diamond surface. Thermal annealing at a temperature of 600 °C in Ar atmosphere was performed in order to improve the ohmic interface between titanium and diamond [14].

A micrograph of the sample is reported in Fig. 2b, together with a sketch of electrical connections' arrangement adopted to perform the IBIC measurements. Also in this case, the current flowing between the two fingers was below our detection limit (i.e. < 1 pA at ±100 V).

IBIC measurements were carried out at the Ruđer Bošković Institute microbeam facility using 4 MeV $He^+$ ions focused to 4 μm diameter spot. The energy of the ion was chosen to probe regions located beneath the graphitic electrodes, as results from the ionization profile shown in Fig. 1.

For the analysis carried out on sample 1, the microbeam raster scanned a rectangular area (120×150 μm$^2$) surrounding the buried electrodes. Ion current was low (<1000 ions·s$^{-1}$) in order to prevent damaging and to avoid pile-up effects. The electronic chain connected to the sensitive electrode comprises a charge sensitive preamplifier ORTEC142 and an ORTEC570 shaping amplifier (shaping time: 0.5 μs). Pulse height processing, beam scanning and 2D map acquisition was carried out by a home-developed hardware and software system [15]. The calibration of the electronic chain was performed using a Si surface barrier detector and a precision pulse generator, in order to relate pulse heights provided by the reference Si detector with those from the diamond device. The spectral sensitivity of the IBIC set-up was ~1100 electrons/channel and the noise threshold level was set to channel 20, corresponding to about 0.4% and 7.2% charge collection efficiency for 4 MeV He ions in diamond, respectively.

A similar experimental set up was adopted for IBIC measurements on sample 2. In this case, the ion beam scan size was about 300×300 μm$^2$.

## 3. Results

Fig. 3a shows the IBIC map collected from sample 1 by the horizontal (sensitive) electrode $S_0$, which is polarized at +80 V; the other vertical electrodes are grounded. Charge pulses overcoming the electronic threshold are generated only in the region surrounding the sensitive electrode; free carriers generated elsewhere do not induce detectable signals. The CCE profile along the horizontal direction centred in the middle of the horizontal electrode is nearly flat, with a small bump at the edge, followed by a rapid fall to zero (Fig. 3b).

When the applied bias is inverted (Fig. 4a), the IBIC map shows a complementary aspect, with no pulses detected in proximity of the sensitive horizontal electrode; detectable pulses are generated in proximity of the four vertical grounded electrodes, as highlighted by the CCE profile (Fig. 4c; solid curve) evaluated along the same direction as in Fig. 3b.

Fig. 4b shows the IBIC map obtained under the following bias configuration: $V_0$=-80 V and the other electrodes with a gradually increasing potential ($V_1$=-60, $V_2$=-40, $V_3$=-20, $V_4$=0 V, respectively). The relevant CCE profiles at the centre of the horizontal electrode is shown in Fig. 4c (dashed curve). In comparison with the case of all the vertical electrodes grounded (IBIC map in Fig. 4a), the profile shows a more pronounced decrease of the maxima as the distance from the sensitive electrode ($S_0$) increases.

The IBIC theory provides a suitable interpretation of these results [11, 16]. If we assume a perfect intrinsic (i.e. no doped) material with ideal ohmic contacts, the traditional approach based on the Ramo's theorem can be adopted, which consists in two different steps: first the weighting potential is mapped by solving the Laplace equation and assuming a unit potential at the sensitive electrode while all the other electrodes are grounded. This allows the identification of the region where charge pulses are generated, whose intensity is proportional to the difference in the weighting potentials between the initial and final position of the moving charges [16]. The second step consists in evaluating the carrier trajectories. This can be ac-complished considering the only drift mechanism of transport (i.e. diffusion is neglected) and solving the Laplace equations with the actual bias potentials at the electrodes.

Figs. 5a, 5b and 5c show the weighting potential maps relevant to the sensitive electrode $S_0$ evaluated by solving numerically the Laplace's equation with the Finite Element Method (FEM) implemented with commercial software (COMSOL Multiphysics 3.5a®) [17]. The streamlines indicate the trajectories of positive charges subjected to the actual electrostatic field generated by the bias configurations relevant to Figs 3a, 4a and 4b, respectively. The generation point was set at a depth of 8 μm below the surface, which corresponds to the maximum of the Bragg ionization curve depicted in Fig. 1.

A comparison of the experimental IBIC maps with the corresponding theoretical potential maps highlights the role of electrons and holes in the charge induction.

In Fig. 3a the signals are generated in proximity of the anode; electrons provide a negligible contribution to the CCE since are suddenly collected by $S_0$ and cross a region with a nearly constant weighting potential. On the other hand, holes are drifted towards the grounded electrodes and span regions with weighting potential ranging from almost 1 to 0 (see Fig. 5a). If the generation occurs nearby the grounded electrode(s), the dominant contribution should be provided by electrons; however the CCE profile in Fig. 3b is null, indicating that the drift length (and thus the lifetime) of electrons is significantly shorter than the distance between the electrodes.

This interpretation is further corroborated by the results shown in Fig. 4a, where the bias polarity has been inverted. In this case, holes generated at the anodes (vertical electrodes) move towards the sensitive electrodes. Their drift lengths are sufficient to span regions with noticeably different values of the weighting potential (Fig. 5b).

It is worth comparing the streamlines in Fig. 5b and 5c. Due to the degrading applied bias potentials at the electrodes, some trajectories generated at $V_2$ converge at $V_1$, spanning almost constant weighting potential values. This is the cause of the difference of the CCE profiles shown in Fig. 4c and relevant to the bias conditions of Figs. 4a and 4b.

Fig. 6a shows the IBIC map and the relevant horizontal profile collected by the second vertical electrode ($S_2$) under the following biasing conditions: $V_0$, $V_2$, $V_4$ grounded and $V_1$, $V_3$ at -100 V. As expected, only the region in proximity of the sensitive electrode (anode) provides a detectable signal induced by the hole transport towards the closer cathodes; the profile is Gaussian-like with a maximum of about 60% occurring at the centre of $V_2$ and a FWHM of about 30 μm. Fig. 5d shows the relevant weighting potential map and hole trajectories.

It is interesting to compare such a profile with the profile obtained in similar bias conditions on sample 2, which was equipped with interdigitated metallic electrodes deposited onto the diamond surface. Fig. 6b shows the relevant IBIC map, which highlights the comb-like structure of the sensitive electrode. The CCE profile shows that at the centre of each tooth-comb, the efficiency reaches a CCE of 80%, a value remarkably higher than what observed in Fig. 6a. The reason of this difference can be attributed to the fact that above-mentioned "cap layer" region has been subjected to a sub-threshold damage induced by ion irradiation. The obtained data indicate that after annealing the diamond structure has not been fully recovered [18,19].

From the ionization curve in Fig.1, it is apparent that about 25% of the carriers are generated in the cap layer. Assuming that in this region the residual damage is still active in inducing efficient hole trapping, we can reasonably expect a reduction of 25% of CCE with respect to the pristine material as supposed to be the sample 1.

## 4. Conclusions

In this paper we report on the IBIC characterization of a CVD diamond ionization radiation detector with buried graphitic electrodes fabricated by DIBL.

From the point of view of the material qualification, IBIC maps acquired under different polarizing bias conditions have evidenced that induced charge signals are formed mainly at the anodes, whatever are the bias polarization conditions and/or location of the sensing electrode. The interpretation of such a fact is based on the assumption that the material is intrinsic and the electrical contacts are ohmic. Such an assumption is supported by the fact that the currents between the electrodes are below the detection limit, whatever the bias polarities are. The experimental results are hence compatible with a model which considers holes as the dominant carrier responsible for the induced charge formation. FEM analysis based on the adjoint equation approach [20,21], not reported here, provide simulations consistent with the experimental results with regards to the negligible value of the electron lifetime (i.e., few ps). The abovementioned FEM analysis assumes hole lifetime values of the order of 0.4 ns in the pristine material, corresponding to maximum drift lengths of the order of few tens of micrometers. These values allow holes to cross the whole active region beneath the graphitic electrodes. From the point of view of the fabrication process qualification, it has been shown that the layer overlying the graphitic channels show traces of a residual damage induced during ion micromachining, which was only partially healed by thermal annealing. Residual trapping centres are therefore present, which strongly reduce the carrier (hole) lifetimes (few ps), making the cap layer (few micrometer beneath the surface) almost inactive for the detection of ionizing radiation.

**Acknowledgements**

This work was supported by INFN experiment DIAMED; by European Community as an Integrating Activity 'Support of Public and Industrial Research Using Ion Beam Technology (SPIRIT)' under EC contract no. 227012; by MIUR, PRIN2008 National Project "Synthetic single crystal diamond dosimeters for application in clinical radiotherapy"; and by University of Torino "Progetti di Ricerca di Ateneo-Compagnia di San Paolo-2011- Linea 1A, progetto ORTO11RRT5".

**References**

[1] P. Olivero, J. Forneris, M. Jakšic, Z. Pastuovic, F. Picollo, N. Skukan, E. Vittone, Nucl. Instr. Meth. B 269 (2011) 2340.

[2] F. Picollo, D. Gatto Monticone, P. Olivero, B. A. Fairchild, S. Rubanov, S. Prawer, E. Vittone, New Journal of Physics 14 (2012) 053011.

[3] M. K. Zalalutdinov et al., Nano Letters 11 (2011) 4304.

[4] P. Olivero et al., Advanced Materials 17 (2005) 2427.

[5] E. Vittone et al., Mater. Res. Soc. Symp. Proc. 1203 (####) 1203-J17-06.

[6] F. Picollo et al., Laboratori Nazionali di Legnaro Annual Report 2011 (2012) 148, ISSN 1828-8545.

[7] S. Prawer et al., Appl. Opt. 34 (1995) 636.

[8] T.I. Galkina, et al., Physics of the Solid State 49 (2007) 654.

[9] C. Davia, S. J. Watts. Nucl. Instr. Meth. A 603 (2009) 319.

[10] M. B. H. Breese, E. Vittone, G. Vizkelethy, P.J. Sellin, Nucl. Instr. Meth. B 264 (2007) 345.

[11] S. Almaviva et al., Nucl. Instr. Meth. A 612 (2010) 580.

[12] J.F. Ziegler, M.D. Ziegler MD, J.P. Biersack, Nucl. Instr. Meth. B, 268 (2010) 1818.

[13] R. Kalish, S. Prawer., Nucl. Instr. Meth. Phys. Res. B 106 (1995) 492.

[14] M. Werner, Semicond. Sci. Technol, 18 (2003) S41-6.

[15] M. Bogovac, M. Jakšić, D. Wegrzynek, A. Markowicz, Nucl. Instr. Meth. B 267 (2009) 2073.

[16] V. Radeka, Ann. Rev. Nucl. Part. Sc. 38 (1988) 217.

[17] COMSOL Multiphysics®, http://www.comsol.com.

[18] P. F. Lai et al., Diam. Relat. Mater. Vol 10 (2001) 82.

[19] E. Baskin et al., Phys. Rev. B 64 (2001) 224110.

[20] E. Vittone, N. Skukan, Z. Pastuović, P. Olivero, M. Jakšić, Nucl. Instr. Meth. B 267 (2009) 2197.

[21] T.H. Prettyman, Nucl. Instr. Meth. A 428 (1999) 72.

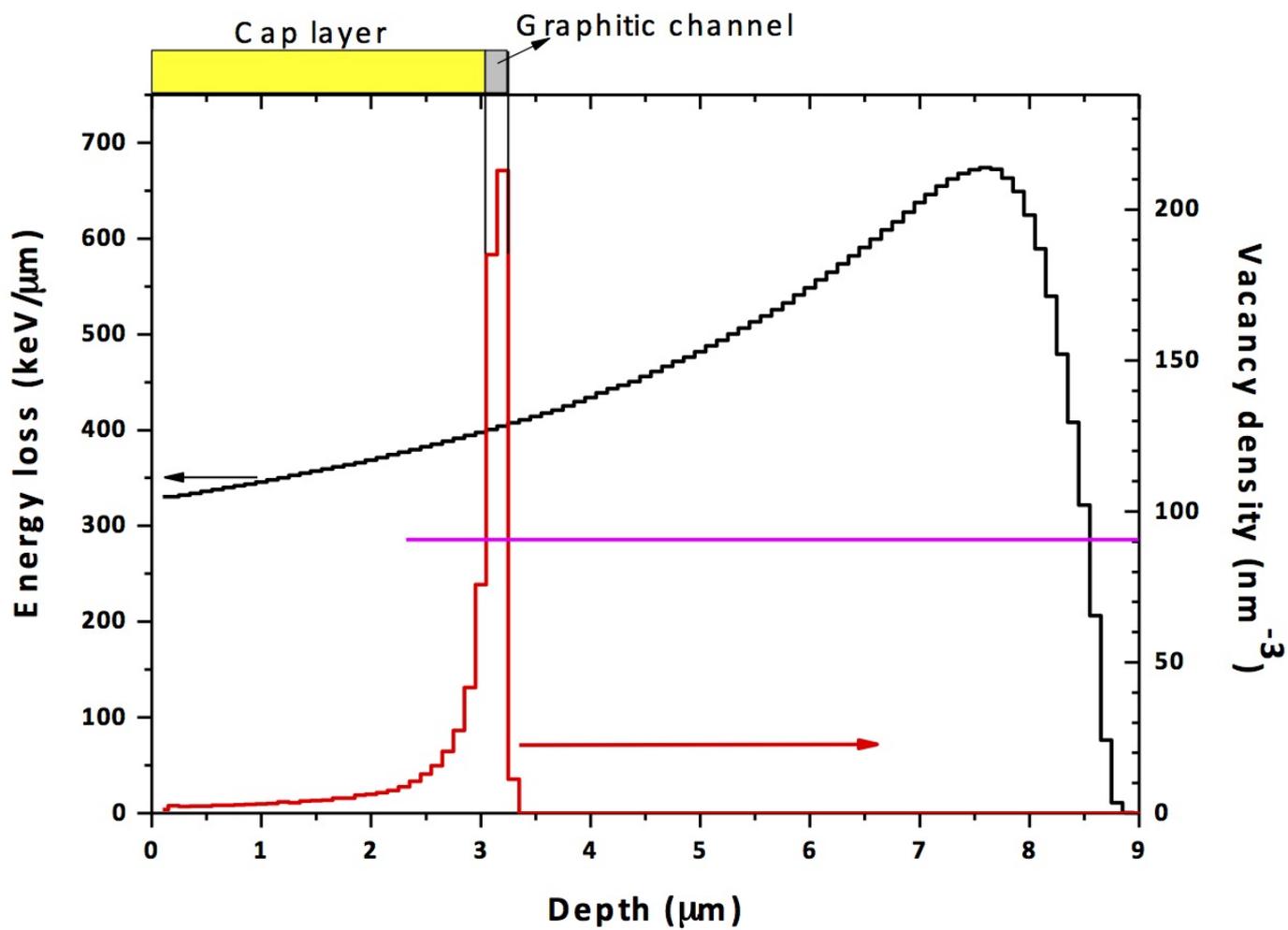

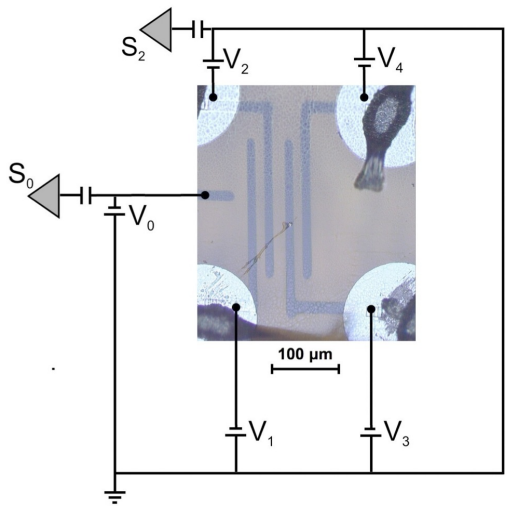 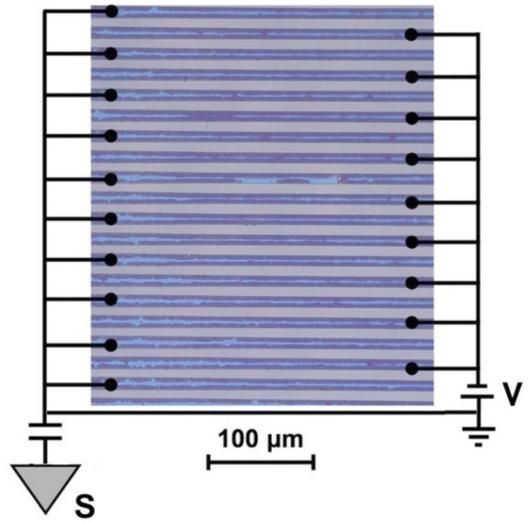

(a) (b)

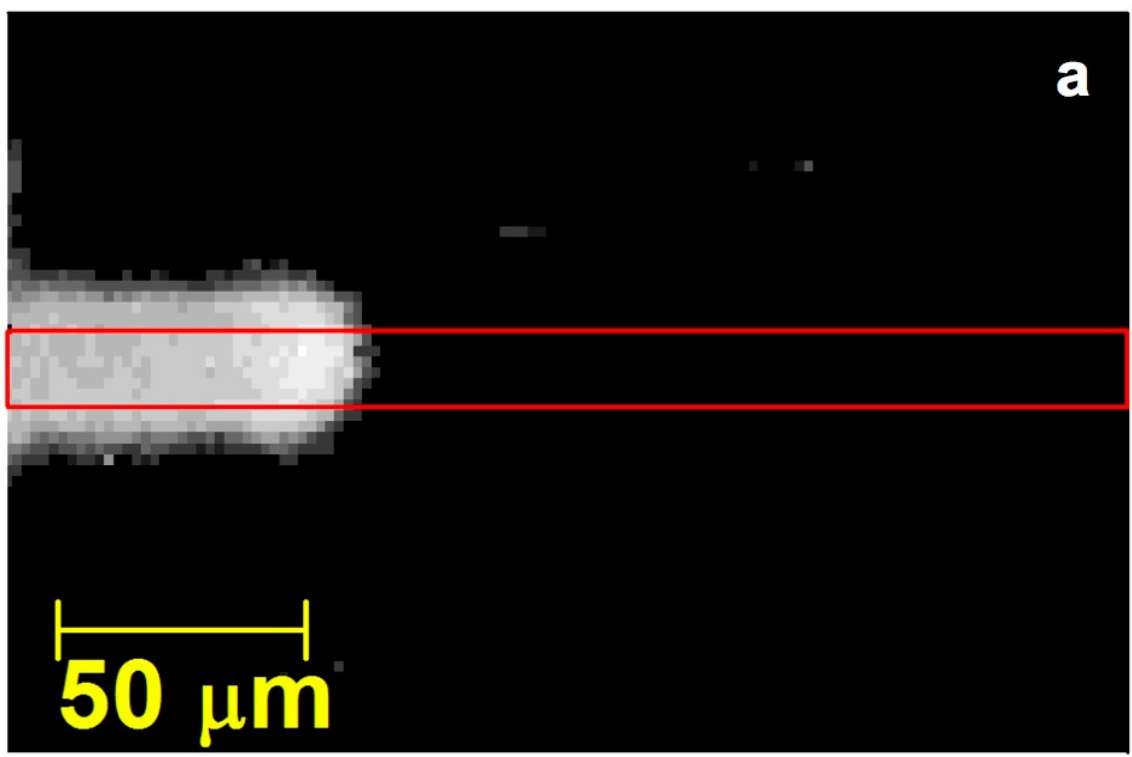
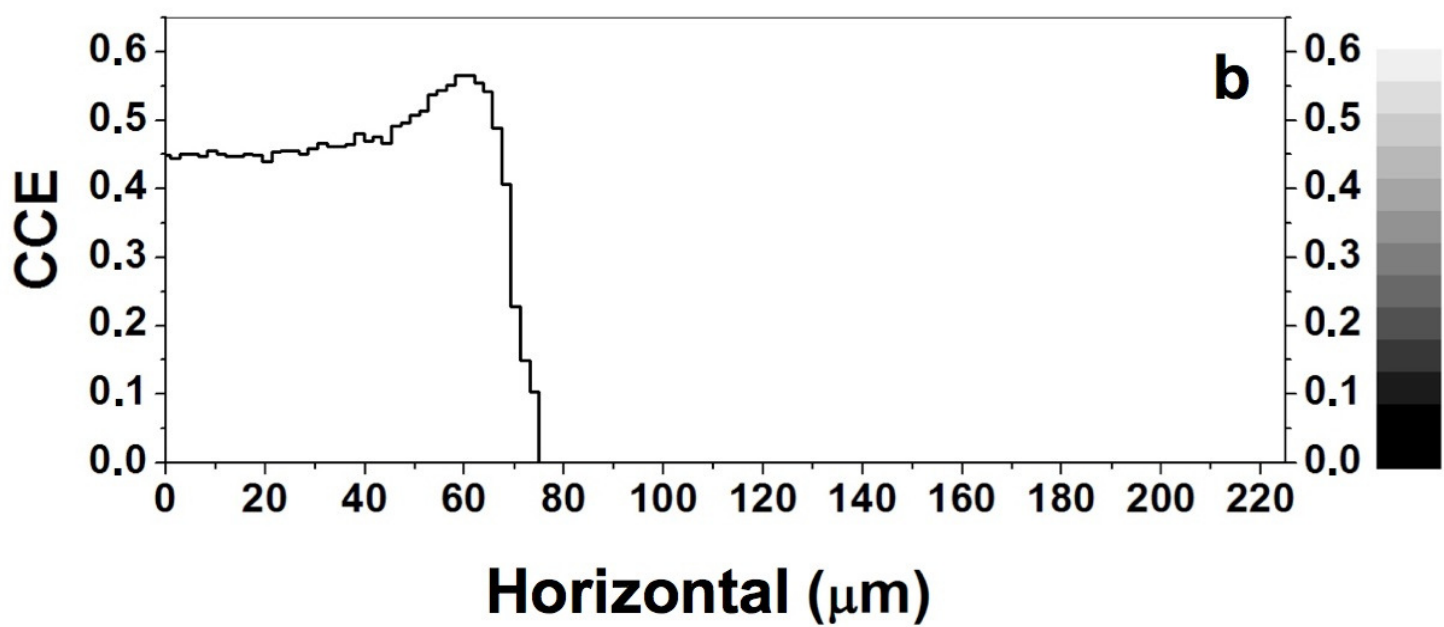

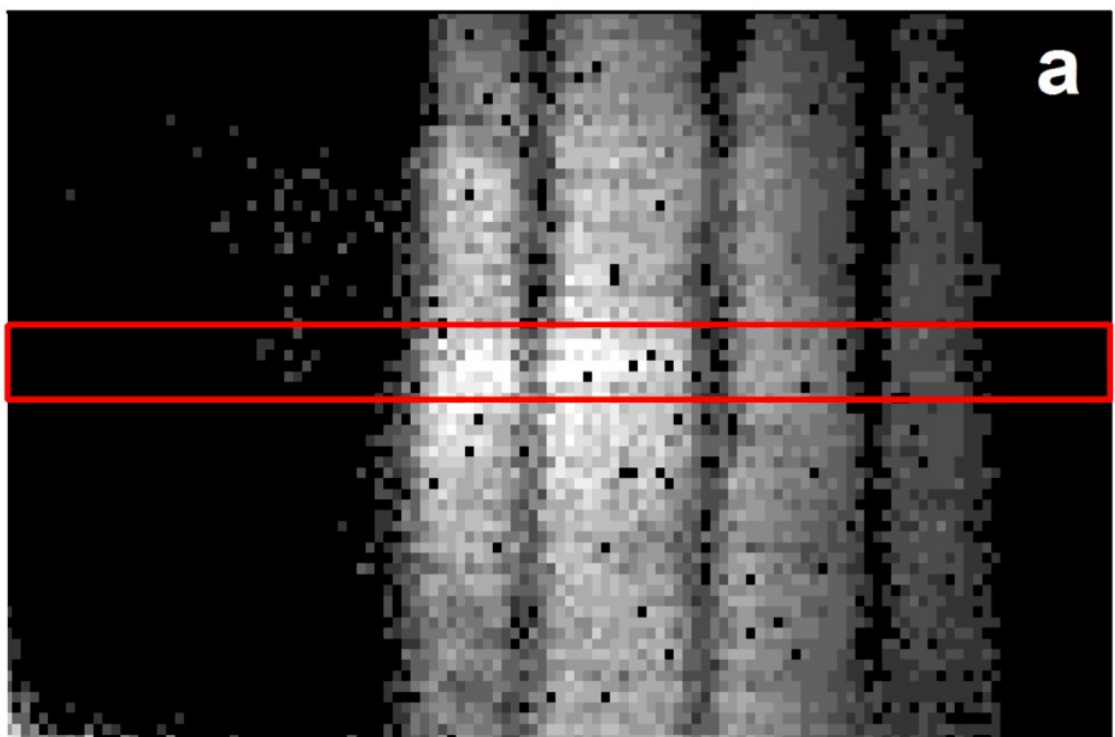
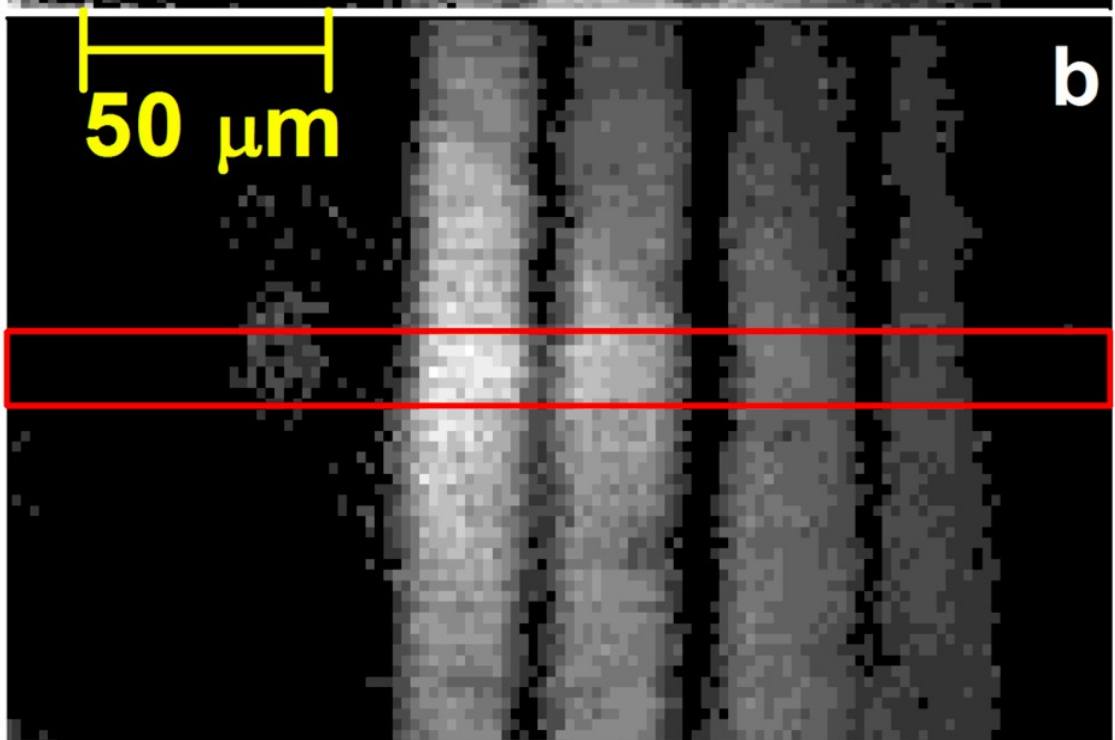
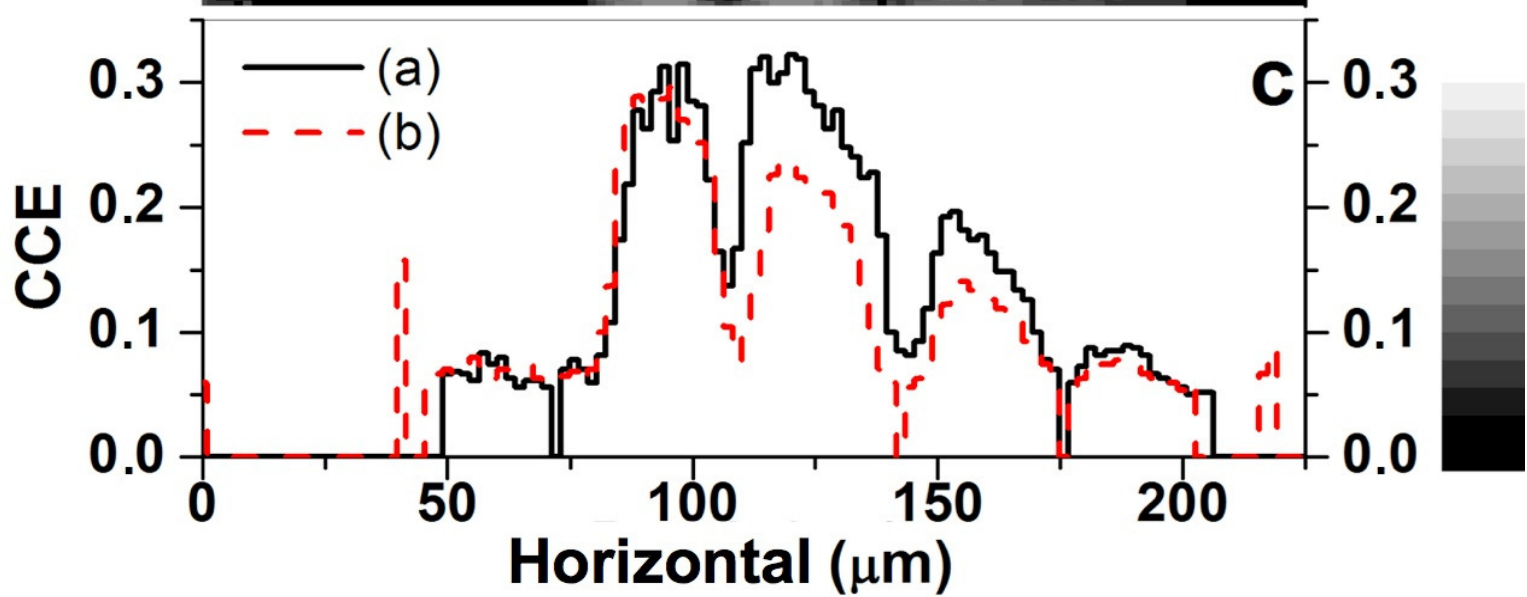

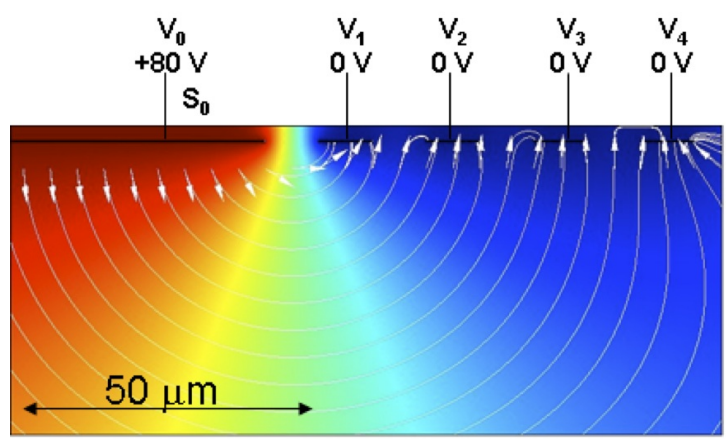
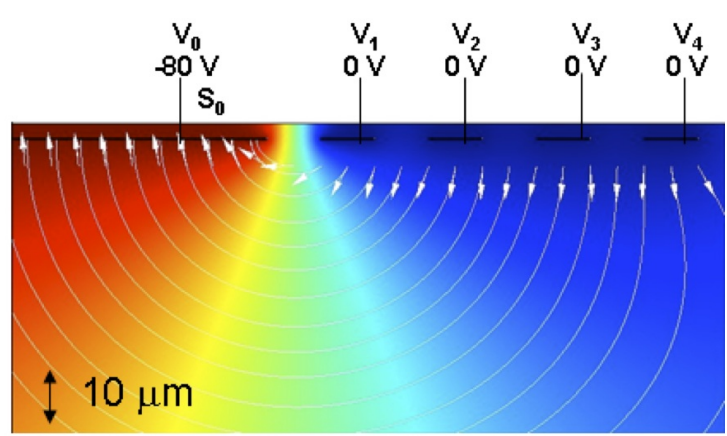
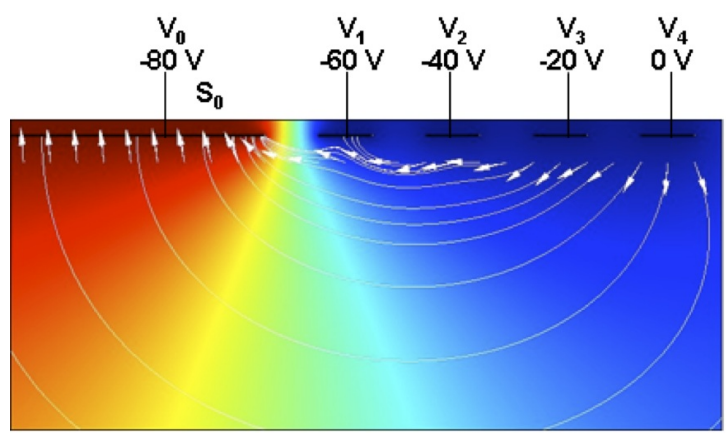
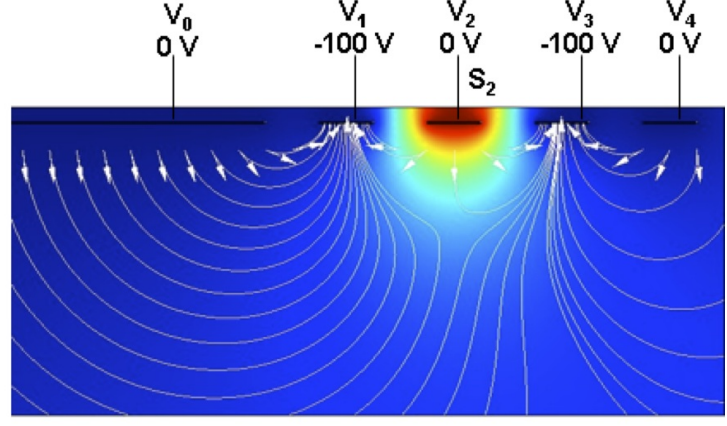

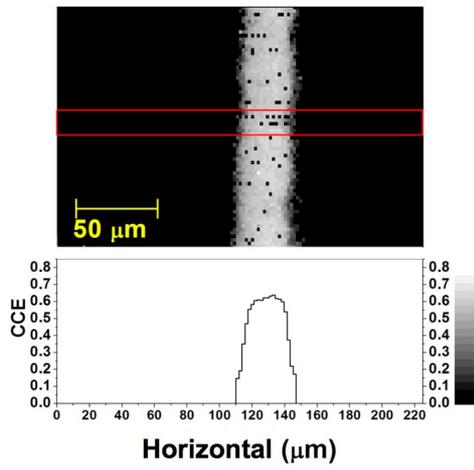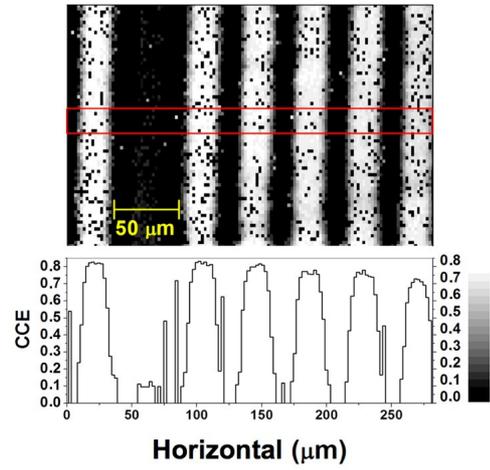

(a) (b)